\documentclass[aps,prl]{revtex4}

\usepackage{bm}
\usepackage{amsmath}
\usepackage{amssymb}
\usepackage{graphicx}
\usepackage{epsfig}
\usepackage{epstopdf}
\usepackage[colorlinks]{hyperref}
\begin{document}

\def\pr{\prime}
\def\be{\begin{equation}}
\def\en#1{\label{#1}\end{equation}}
\def\d{\dagger}
\def\bar#1{\overline #1}
\newcommand{\per}{\mathrm{per}}
\newcommand{\m}{\mathbf{m}}
\newcommand{\n}{\mathbf{n}}
\newcommand{\bb}{\mathbf{b}}
\newcommand{\ab}{\mathbf{a}}
\newcommand{\zb}{\mathbf{z}}
\newcommand{\alphab}{\boldsymbol{\alpha}}
\newcommand{\betab}{\boldsymbol{\beta}}

\newcommand{\rd}{\mathrm{d}}
\newcommand{\vare}{\varepsilon }
\def\NK #1{\textcolor{blue}{#1}}

\title{Trading quantum states for temporal profiles: tomography by the overlap}
\author{J. Tiedau$^1$, V.~S.~Shchesnovich$^2$, D.~Mogilevtsev$^{2,3}$, V.~Ansari$^1$, G. Harder$^1$, T.~Bartley$^1$, N.~Korolkova$^4$ and Ch.~Silberhorn$^{1}$}

\affiliation{
$^1$Applied Physics, University of Paderborn, Warburgerstrasse 100, 33098 Paderborn, Germany;
\\
$^2$Centro de Ciencias Naturais e Humanas, Universidade Federal do
ABC, Santo Andre,  SP, 09210-170 Brazil;
\\
$^3$Institute of Physics, Belarus National Academy
of Sciences, F.Skarina Ave. 68, Minsk 220072 Belarus;
\\
$^4$School of Physics and Astronomy, University of St Andrews,
North Haugh, St Andrews KY16 9SS, UK}

\begin{abstract}
Quantum states and the modes of the optical field they occupy are intrinsically connected. Here, we show that one can trade the knowledge of a quantum state to gain information about the underlying mode structure and, vice versa,
the knowledge about the modal shape allows one  to perform a complete tomography
of the quantum state. Our scheme  can be  executed  experimentally using the  interference between the  signal and probe states on  an unbalanced beam splitter with a single on/off-type  detector. By changing the  temporal overlap between the  signal and the probe,  the  imperfect interference is turned into a powerful tool to extract the information about the signal mode structure. A single on/off detector is already sufficient to  collect the necessary  measurement data for the  reconstruction of  the diagonal part of the density matrix of an arbitrary multi-mode signal. Moreover, we experimentally demonstrate the  feasibility of our scheme with just one control parameter --  the   time-delay of a coherent  probe  field.

\end{abstract}

\pacs{14 August 2017}
\maketitle

\section{I. Introduction}
Encoding quantum states in modes of optical signals is the basis of optical quantum information, communication and metrology. A specific example is the use of spectral-temporal modes of light.
The art of temporal mode shaping for use in quantum information is already quite advanced~\cite{eckstein,brecht,raymer}, and for example can be used for enhancing the resolution in spectroscopy~\cite{mukamel}.
One can transform one pulse shape into another preserving its quantum state and single-mode character,  for example, using a quantum pulse gate~\cite{eckstein,brechtnjp,reddy,manur}.

Extraction of information on  the quantum state encoded into modes of light is the subject of  the quantum state reconstruction, in practice performed  by mixing the signal with some known reference modes.
When the modal profiles and structures of the signal and probes are not the same, this common  approach for the state reconstruction becomes increasingly challenging and impractical. Furthermore, it is not  the mere increase in required number of copies that renders the scheme impractical  for reaching a prescribed accuracy; the problem can be far deeper.

Let us consider, for example, a standard technique of quantum homodyne tomography. The signal state is optically amplified by a bright local oscillator, with information extracted from the noise of the resulting interference pattern. In homodyne detection, the bright local oscillator acts as a perfect mode filter: only information present in the same mode profile as that of the local oscillator is extracted; all other information is lost. Therefore, both precise knowledge of the mode structure of the signal and precise manipulation of the mode of the local oscillator are required.  Here the low overlap between the signal and bright probe is equivalent to high loss and might render the reconstruction completely unfeasible~\cite{vogel0,all}. Imperfect overlap between the signal and the probe is  generally believed to be the main  hindrance  for any reconstruction  scheme relying on the  interference. Indeed, an imperfect overlap---typically referred to as ``distinguishability'' of modes ---degrades visibility of interference fringes. Distinguishability of the spatiotemporal modes is also highly detrimental to quantum devices  that require for their operation many single photons with  identical spatiotemporal profiles, such as Boson Sampling~\cite{aaron}, where scalability to higher photon numbers imposes severe restrictions on the overlap of the single photons at the input~\cite{valera1}.

In this paper, we argue that the very same distinguishability can be used in a constructive way in
the same spirit as noise and losses were exploited in
lossy on/off tomography and noise tomography \cite{mog98,us,paris,mogilevtsev2009,harder}.
We propose a new type of state reconstruction scheme exploiting imperfections. This scheme as well -- contrary to standard homodyne  detection -- requires probe and signal of comparable intensities. Recently, non-balanced detection schemes (or homodyning with weak probe) are being extensively studied both theoretically and experimentally (see, for example, Refs. \cite{1,2,3,4,5}). 
The distinctive feature of our scheme is that it is in fact based on distinguishability, thus turning a hindrance to the state reconstruction into an ally.  Such scheme delivers all the information sufficient for the state reconstruction and this is due to some of its intrinsic features explained below. The measured data resulting from the signal-probe interference depends on both the quantum state and the mode overlap between the interfering modes. Importantly, the scheme is sensitive to the whole signal, both parts overlapping with the probe and non-overlapping parts. This is a fundamental trade-off feature of our scheme: one can trade knowledge about a quantum state for its  modal profile, and  vice versa. {Furthermore}, it can be done with the same measurement set-up, and from this data, one can determine the fraction of the quantum state that overlaps with the local oscillator, in stark contrast to the standard homodyne detection. Moreover, a non-balanced detection scheme can be used to even extract  information on the quantum state of a signal without prior  knowledge on its underlying mode structure \cite{mogilevtsev2009}.

The scheme proposed in this work  relies on the interference of the unknown signal field and a reference field of known temporal profile and in a known quantum state. In order to  collect data allowing for either complete reconstruction of the signal state or of its temporal profile, there is no need to vary
the temporal profile of the reference field. It is sufficient to delay
the reference pulse in a controlled way. We show that
with the reference field described by a diagonal density matrix it is still possible to infer both the temporal profile and the photon-number distribution of the signal because because the absolute value of the overlap between the signal and the reference fields can be controlled solely by the relative time-delay of the reference.

We present an experimental confirmation of the above with the signal represented by single and two-photon quantum states generated by the spontaneous parametric down-conversion process and with  a single on/off detector.
Moreover, it is remarkable that our simple detection scheme is able to provide  the complete inference of an arbitrary  multi-mode quantum state, provided that one can have available a controlled set of multi-mode probes, each in a quantum coherent state. Thus, we show that the simplest on/off detector is a device able to translate a complete information about the arbitrary quantum state into a sequence of binary signals.

The outline of the paper is as follows. In Sec. II we present the theoretical basis  for our  scheme by describing  how the  multi-mode probe and signal fields interfere on a beam-splitter. Then, we focus on  the particular case of mixing a multimode  signal with the  probes in coherent states to show the possibility to infer the complete signal state.  In Sec. III,   the reconstruction scheme is  elaborated  in detail for the case of   the single-mode signal and probe fields, illustrating the implementation of the data pattern approach for the signal state reconstruction. In Sec. IV we discuss how one can  infer  the  overlap and the temporal/spectral profile of the signal. In particular, we show that for the signal/probe with a diagonal density matrix just a single probe state is sufficient to infer the overlap.  The experimental results are presented  in Sec. V. We have experimentally reconstructed the quantum states of  the single and two-photon signals, and then demonstrated the overlap inference from the knowledge of the quantum state.

\begin{figure}[htb]

\includegraphics[width=0.5\linewidth]{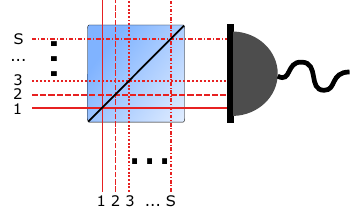}

\caption{ A scheme of the simplest measurement set-up. Two sets of $S$ incoming modes (signal and probe) are combined on a beam splitter. {A set outgoing from one of the beam splitter outputs is measured with an on/off detector.}}

\label{fig1}
\end{figure}

\section{II. The scheme}
First of all, let us show how the state of an arbitrary multi-mode field can be inferred with the simplest measurement set-up
with on/off detectors.
The signal field is mixed with the probe field
on the beam-splitter;  the outputs (or just one of them) are impinging on the bucket, on/off type
detectors as depicted in Fig.~\ref{fig1}.

As illustrated in {Fig.~\ref{fig1}}, the imperfect overlap on the beam splitter (BS) can be modeled by separating the input fields into the modes perfectly overlapping on the BS (``modes mixed by BS'') and the modes with zero overlap thus not interacting on the BS (``modes not mixed by BS''). Then let the two inputs and two outputs  on a beam splitter  be given by the mode creation operators   $a^\dag_{k,s}$ and $b^\dag_{k,s} $, where $k = 1,2$ is the operating mode mixed by the BS {(solid red lines in Fig.~\ref{fig1})}, whereas  $s=1,\ldots,S$ enumerates the basis of the  invariant modes, i.e.,  not mixed  by the BS {(various dashed and dash-dotted lines in Fig.~\ref{fig1})}. The beam splitter itself is represented by a unitary matrix $U$ which relate  the  output to the input operators:

\be
b^\dag_{k,s} = \sum_{l=1,2} U_{k,l} a^\dag_{l,s}.
\en{E1}
In the matrix form  it can be written as $ \bb^\dag  = (U\otimes I)\ab^\dag$, where $I$ is the $S$-dimensional identity matrix and $\ab = (\ab_1,\ab_2)$ with  $\ab_k = (a_{k,1},\ldots,a_{k,S})$, etc.

Consider  an arbitrary multi-mode state at input $k$ and the non-ideal on-off type detector placed at the outputs of the beam splitter. The quantum state $\rho^{(k)}$ of the input modes $k$ can be conveniently written in Fock basis as
\be
\rho^{(k)}= \sum_{\n,\m\ge 0} \rho^{(k)}_{\n,\m} |\n\rangle \langle \m |, \quad |\n\rangle \equiv \prod_{s=1}^S\frac{\left( a^\dag_{ks}\right)^{n_s} }{\sqrt{n_s!}}|0\rangle.
\en{E2}
Note that {in this expression} the multi-mode character of the field is encapsulated in quantities in bold, e.~g., $|\n\rangle$ and $\n\equiv (n_1,\ldots,n_S)$.
Let $\eta_k$ be the sensitivity of the detector at output mode $k=1,~2$. We are interested in the probability that the detector  $k$ does not click. Such probabilities of no clicks at the outputs {(zero-click probabilities)} are given by the generating function:
\begin{eqnarray}
P(\eta_1,\eta_2) &\equiv&  \mathrm{Tr}\left(\rho^{(1)}\otimes\rho^{(2)} \mathcal{N}\exp\left[ - \sum_{k=1,2}\eta_k\sum_{s=1}^S b^\dag_{k,s}b_{k,s}\right] \right) =   \mathrm{Tr}\left(\rho^{(1)}\otimes\rho^{(2)} \mathcal{N}\exp\left\{ - \bb\left[\left(\begin{array}{cc} \eta_1 & 0 \\ 0 & \eta_2\end{array} \right)\otimes I\right] \bb^\dag \right\} \right)\nonumber\\
&=& \mathrm{Tr}\left(\rho^{(1)}\otimes\rho^{(2)} \mathcal{N}\exp\left\{ - \ab\left[U^\dag\left(\begin{array}{cc} \eta_1 & 0 \\ 0 & \eta_2\end{array} \right)U\otimes I\right] \ab^\dag \right\} \right)
\label{E3}\end{eqnarray}
where $\mathcal{N}$ denotes the normal ordering of the creation and annihilation operators. For example, to calculate the probability of no clicks at detector $1$, $p_1(\eta_1)$, we set $\eta_2=0$ in Eq.~(\ref{E3}), that is $p_1(\eta_1) = P(\eta_1,0)$.

To compute the probabilies from the family of $P(\eta_1,\eta_2)$ in Eq.~(\ref{E3}), we use  the following  identity
\be
\mathcal{N} \exp \left\{ -\ab( I - H)\ab^\dag  \right\} = \frac{\mathcal{A}\exp\left\{ -\ab( H^{-1}-I )\ab^\dag   \right\}}{\mathrm{det}(H)},
\en{E4}
where   $\mathcal{A}$ denotes the antinormal ordering of the creation and annihilation operators and $H$ is some positive semi-definite Hermitian matrix with  eigenvalues bounded by $1$.  Eq. (\ref{E4}) can be verified by diagonalizing the matrix $H$, performing the series expansion of the exponents and checking  the equality of both  sides  using the Fock states. In Eq. (\ref{E3}) we can identify the matrix $H$ and find its inverse as:
\be
H  = U^\dag \left(\begin{array}{cc} 1- \eta_1 & 0 \\ 0 & 1-\eta_2\end{array} \right)U\otimes I,\quad H^{-1} =  A\otimes I, \quad A\equiv U^\dag \left(\begin{array}{cc} \frac{1}{1- \eta_1} & 0 \\ 0 & \frac{1}{1-\eta_2}\end{array} \right) U.
\en{E5}
Note also that  $\mathrm{det}(H) = (1-\eta_1)^S(1-\eta_2)^S$.
Substituting Eqs.~(\ref{E4}) and (\ref{E5}) into Eq.~(\ref{E3}) we then obtain:
\be
P(\eta_1,\eta_2) = \frac{1}{ (1-\eta_1)^S(1-\eta_2)^S} \mathrm{Tr}\left(\rho^{(1)}\otimes\rho^{(2)} \mathcal{A}\exp\left\{ -\ab([A-I]\otimes I )\ab^\dag   \right\} \right).
\en{E6}
We can evaluate the trace in Eq. (\ref{E6}) in the coherent state basis $|\alphab_k\rangle\equiv |\alpha_{k,1}\rangle\otimes\ldots\otimes |\alpha_{k,S}\rangle$ by introducing the generalized (multimode) Husimi function for each input state,
\be
Q^{(k)}(\alphab,\alphab^\dag) \equiv \frac{1}{\pi^S} \langle \alphab|\rho^{(k)}|\alphab\rangle.
\en{E7}
The generalized Hiusimi function is
normalized, $\int d\mu(\alphab) Q^{(k)}(\alphab,\alphab^\dag) = 1$ where $d\mu(\alphab) \equiv \prod_{s=1}^Sd^2\alpha_s$ with $d^2\alpha_s = d\mathrm{Re}\alpha_s d\mathrm{Im}\alpha_s$ and $\alphab = (\alphab_1,\ldots,\alphab_S)$. From Eqs. (\ref{E6}) and (\ref{E7}) we obtain
\be
P(\eta_1,\eta_2) = \int d\mu(\alphab_1)\int d\mu(\alphab_2) Q^{(1)}(\alphab_1,\alphab^\dag_1) Q^{(2)}(\alphab_2,\alphab^\dag_2) \frac{\exp\left\{ -\vec{\alphab} \left([A-I]\otimes I \right)\vec{\alphab}^\dag \right\}}{(1-\eta_1)^S(1-\eta_2)^S},
\en{E8}
Here we have introduced the combined vector variable $\vec{\alphab} \equiv (\alphab_1,\alphab_2)$ with $\alphab_k = (\alpha_{k,1},\ldots,\alpha_{k,S})$. The probabilities of no click given by Eq. (\ref{E8}) provide the key relation for the reconstruction of an arbitrary multimode field in quantum state $\rho^{(1)}$  by mixing it with a probe field $\rho^{(2)}$ on a beam splitter with the use of the non-ideal on-off type detector(s).

For a set of coherent probes available in different modes the scheme becomes particularly simple. It allows us to infer the multi-mode signal collecting a string of zeros and ones on merely one on/off detector. Let $\rho^{(2)} = |\betab\rangle\langle \betab|$ for some $\betab = (\betab_1,\ldots,\betab_S)$.  We have
\be
Q^{(2)}(\alphab_2,\alphab^\dag_2) = \frac{1}{\pi^S} \exp\left\{-|\alphab_2- \betab|^2 \right\}.
\en{E9}
To derive the expression for the probabilities, we use the identities for the BS transformation matrices: $|U_{k1}|^2 + |U_{k2}|^2 = 1$, $U_{11}U^*_{21} = - U_{12}U^*_{22}$, $|U_{11}| = |U_{22}|$, and $|U_{12}| = |U_{21}|$, which result from the unitarity  of the $2$-dimensional matrix $U$ of the beam splitter. For the Gaussian integrals, the following standard relation holds:
\be
\int \frac{d\mu(\alphab)}{\pi^S} \exp\left\{ - \alphab B \alphab^\dag + \alphab \mathbf{h}^\dag + \mathbf{g}\alphab^\dag\right\}  = \frac{\exp\{\mathbf{g} B^{-1}\mathbf{h}^\dag\}}{\mathrm{det}(B)},
\en{E10}
where $\alphab = (\alpha_1,\ldots,\alpha_S)$, $B$ is a positive semi-definite Hermitian matrix, while $\mathbf{g}$ and $\mathbf{h}$ are  $S$-dimensional vector rows.
Substituting (\ref{E9}) into (\ref{E8}) and evaluating the  Gaussian integral  over $\alphab_2$ using Eq.~(\ref{E10}), we obtain
\be
P(\eta_1,\eta_2) = \varkappa^{-S}\exp\left\{- \frac{\eta_1\eta_2}{1-\varkappa}|\betab|^2 \right\}\int d\mu(\alphab)\exp\left\{  -\frac{1-\varkappa}{\varkappa} \left|\alphab+ \frac{(\eta_1-\eta_2)U_{11}U^*_{21}}{1-\varkappa}\betab\right|^2\right\}Q^{(1)}(\alphab,\alphab^\dag)
\en{E11}
where $ \varkappa = 1- \eta_1|U_{11}|^2-\eta_2|U_{21}|^2$.

Now, notably, for the coherent probes $\rho^{(2)} = |\betab\rangle\langle \betab|$, the  reconstruction of the multi-mode source field $\rho^{(1)}$ is possible from the no clicks probability even on just one detector, $P(\eta_1, 0)$, setting  $\eta_2=0$ in Eq. (\ref{E11})). This is due to  the fact that  the expression in Eq. (\ref{E11}) is proportional to a multidimensional  Weierstrass transform of the  Husimi function of the source field, which allows for  the inverse transform by the properties of the Husimi function (see Appendix).

\section{III Example:  a single-mode  source field }

We aim at the demonstration of trading the information about the spectral state properties for the knowledge of the quantum state and vice versa. Now let us consider a methodologically and practically important case of the probe and signal fields in single-mode spectrally pure states.
We assume that the input temporal modes (TM) of the same polarization in {Fig.\ref{fig1}} are described by the collective operators \be A_{x}=\int\limits_0^{\infty}d\omega
f_x(\omega)a_x(\omega), \en{definition1} where the index $x$
denote{s} either signal, $s$, or probe, $p$, modes; $a_{x}(w)$ is the
annihilation operator of the photon in the plane wave with the
frequency $\omega$; $\int\limits_0^{\infty}d\omega
|f_x(\omega)|^2=1$. Again, we assume that both inputs are
{single mode} and in the same spatial mode.  Thus, a degree of distinguishabity is described by the overlap \be
\phi_1=\int\limits_0^{\infty}d\omega f_s(\omega)f_p(\omega)^*.
\en{overlap} We assume that the BS acts similarly on input plane wave modes of
arbitrary frequency; the BS action is described by the unitary
$2\times2$ matrix $U$. Initial states are described by the density
matrices $\rho_x$ depending only on the corresponding collective
mode operators.

{To apply the formalism developed in the previous section for our single-mode case, we need to add auxiliary orthogonal vacuum modes to both signal and probe inputs. Thus, the two-mode internal basis  (i.e., $S=2$) is used in this case.}
The operator of the original source field mode, $c^\dag_1$,  is expressed through the internal basis operators as $c^\dag_1\equiv \phi_1 a^\dag_{1,1} + \phi_2 a^\dag_{1,2}$ with $|\phi_1|^2+ |\phi_2|^2 = 1$. We introduce the auxiliary mode $c^\dag_2$ orthogonal to $c^\dag_1$,  and write down the operators of the internal basis, $\ab_1 = (a_{1,1},a_{1,2})$ as
\be
\ab^\dag_1= V\mathbf{c}^\dag, \quad V = \left(\begin{array}{cc} \phi_1^* & -\phi_2 \\ \phi_2^* & \phi_1 \end{array} \right),
\en{E13}
where $\mathbf{c} = (c_1,c_2)$.
The single-mode source field reads:
\be
{\rho}^{(1)} = \sum_{n,m\ge 0} \rho^{(1)}_{n,m} \frac{(c^\dag_1)^n}{\sqrt{n!}}|0\rangle\langle 0|\frac{(c_1)^m}{\sqrt{n!}}
\en{E12}
{(note that the basis $c_{1,2}$ is used for the expansion of the state).}
Eq. (\ref{E13}) leads to an analogous  relation between the parameters  of the respective   coherent states via the identity $\alphab_1 \ab_1^\dag = \alphab_1 V \mathbf{c}^\dag \equiv \alphab \mathbf{c}^\dag$, i.e.,  the basis coherent vector $\alphab_1 = (\alpha_{1,1},\alpha_{1,2})$ in the $\ab$-basis is related to the  coherent vector $\alphab=(\alpha,\alpha^\prime)$ in the $\mathbf{c}$-basis. In the latter representation,  the Husimi function  of the above  single-mode source  (\ref{E12}) becomes
\be
Q^{(1)}(\alphab_1,\alphab^\dag_1) = \frac{e^{-|\alpha^\prime|^2}}{\pi} Q(\alpha,\alpha^*), \quad Q(\alpha,\alpha^*) \equiv \frac{1}{\pi}\langle \alpha|{\rho}^{(1)}|\alpha\rangle,
\en{E14}
where $|\alpha\rangle$ is the shortcut notation for the state $|\alpha,0\rangle$ in the new basis $c_{1,2}$ of Eq. (\ref{E13}).

The integration measure $d\mu(...)$ is invariant under the  unitary transformation in Eq. (\ref{E13}). We need to evaluate the Gaussian integral over the variable $\alpha^\prime$ in Eq. (\ref{E11}). We need only one detector for reconstruction, select the $k$th detector.  We then obtain the the probability of no clicks at detector $k$ when mixing the single-mode source field ${\rho}^{(1)}$ (\ref{E12}) with the single-mode coherent probe in the $\ab$-basis $\rho^{(2)} = |\beta\rangle\langle\beta|\otimes|0\rangle\langle0|$ (i.e., $\betab = (\beta,0)$ in Eq. (\ref{E11})):
\begin{eqnarray}
\label{E91}
&& p_k = \frac{e^{-\gamma_k}}{1-\eta_k|U_{k,1}|^2}\int\rd^2\alpha\,e^{ -\sigma_k|\alpha - z_k|^2 } Q(\alpha,\alpha^*),\nonumber\\
&& \gamma_k = \eta_k|U_{k,2}|^2 |\phi_2|^2|\beta|^2,\quad \sigma_k = \frac{\eta_k|U_{k,1}|^2}{1-\eta_k|U_{k,1}|^2},\quad z_k = -\left(\frac{U_{k,2}}{U_{k,1}}\phi_1\right)^*\beta.
\end{eqnarray}

Eq.(\ref{E91}) is a generalization for the imperfect overlap of the result obtained in Ref.\cite{vogel0} for the ideal overlap. As we shall see, this equation points directly to the possibility
of the signal state reconstruction.

To infer a set of discrete parameters, such as density matrix coefficients in the Fock states basis,
it is not necessary to revert to the inverse Weierstrass transform. In general, one can reconstruct an unknown
state of the signal mode by taking a discrete set of the amplitudes, $\beta$, of the
coherent probe for a fixed overlap, $\phi_1$, between the modes
(which can be equal to 1 as a special case). However,
with any given accuracy, one can also reconstruct an
unknown state just by varying the overlap for a fixed amplitude
$\beta$ of the coherent mode. To that end, let us consider
representation of $\rho_s$ in some discrete basis. For example, in Fock-state basis the $Q$-function can be represented as \be Q(\alpha,\alpha^*) =
\frac{1}{\pi}e^{-|\alpha|^2}\sum_{n,m\ge 0}
\rho_{n,m}\frac{(\alpha^*)^n\alpha^m}{\sqrt{n!m!}}, \en{E121} where
the matrix elements $\rho_{n,m} \equiv \langle n|
\rho_s|m\rangle$. From Eqs. (\ref{E91}-\ref{E121}) it
straightforwardly follows that the probability of zero clicks as a function of parameter $z$  is given by:
\be p(z,z^*) =
\exp\{-\eta_k|U_{k,2}|^2|\beta|^2\}\mathcal{P}(z,z^*),
\en{E131}
where $\mathcal{P}(z,z^*)$ is an expansion in \textit{non-negative
powers} of $z$ and $z^*$, which reads
\begin{eqnarray}
 \mathcal{P}(z,z^*) = \sum_{n,m\ge 0}
\frac{\rho_{n,m}}{\sqrt{n!m!}} \exp\left(
-\frac{|\mu|^2}{1+\sigma}\right)
\partial^n_\mu
\partial^m_{\mu^*}\exp\left(
\frac{|\mu|^2}{1+\sigma}\right)\biggr|_{\mu = \sigma z}.
\label{E141}
\end{eqnarray}
Eq.~(\ref{E131}) points to some important conclusions. First of all,
for the realistic finite reconstruction subspace (corresponding to
the truncation of the Fock state expansion in (\ref{E12})), one can
indeed infer the signal state for a finite number of $z$ values.
It is  always possible to choose such a number of $z$ values
as to interpolate a continuous function $\mathcal{P}(z,z^*)$ by the Lagrange polynomials with sufficient accuracy. This allows then to infer the
$Q$-function. Of course, it is more practical to infer a discrete
set of expansion coefficients, $\rho_{m,n}$. This can be done in a
number of already well-established ways (for example, by inverting
Eq.~(\ref{E13}) using constrained least-square methods 
\cite{data-patterns2013,motka}, or maximum-likelihood methods
\cite{mogilevtsev2009}).
From Eq.~(\ref{E13}) it also follows that one can get the set of
$z$-points sufficient for reconstruction of any finite-subspace
expansion of $\rho_s$ for an arbitrarily small probe state
amplitude, $\beta$.

To realize the scheme in practice, it is necessary to devise a way to change the
overlap to produce the set of values, $\{z_j\}$, sufficient for
the inference. Nowadays there are quite sophisticated methods for
shaping the spectral profile of the mode (see, for example,
Refs.\cite{bellini,perer}). However,
one really does not need to manipulate precisely a shape of the
probe pulse to obtain the necessary set; a simple time-delay
arrangement is sufficient.

For the inference it is illustrative to
represent an arbitrary signal as the finite sum of the coherent
projectors, \be
\rho_s=\sum\limits_jc_j|\alpha_j\rangle\langle\alpha_j|,
\en{pattern} where the coefficients $c_j$ are real, but not
necessarily positive. Such a representation is usually implemented
in data-pattern tomography schemes \cite{data-patterns2013,motka,
our-prl2010,oxford,paderborn}. Thus, for different
values of the overlap, $\phi_{1,l}$, the no-click probability is
given by the following expression:
\begin{eqnarray}
 p_l=
\exp\{-\eta|U_{1,2}|^2|\phi_{2,l}|^2|\beta|^2\}\langle{\bar\beta}_l|\bar{\rho}_s|{\bar\beta}_l\rangle,
 \label{represent}
\end{eqnarray}
where
${\bar\rho}_s=\sum\limits_jc_j|\bar{\alpha}_j\rangle\langle\bar{\alpha}_j|$,
$\bar{\alpha}_j=\sqrt{\eta}|U_{1,1}|\alpha_j$,
and
${\bar\beta}_l=-\sqrt{\eta}\exp\{i\arg(U_{1,1})\}U_{1,2}^*\phi_{1,l}^*\beta$. {That is, the density matrix ${\bar\rho}_s$ in (\ref{represent})  is represented by coherent projectors sitting in the smaller region on the phase plane than in case of the original matrix~(\ref{pattern})}.
Note that the matrix ${\bar\rho}_s$ can be arbitrary as well.
Eq.(\ref{represent}) links the probability of no clicks with the signal density matrix in terms of coherent projectors (\ref{pattern}). That means, if it is possible to represent accurately the signal in
terms of the coherent state projectors with the amplitudes
${\bar\beta}_l$, then our scheme allows for the reconstruction of the coefficients $c_j$ in Eq.(\ref{pattern}).

\begin{figure}[htb]
\includegraphics[width=0.75\linewidth]{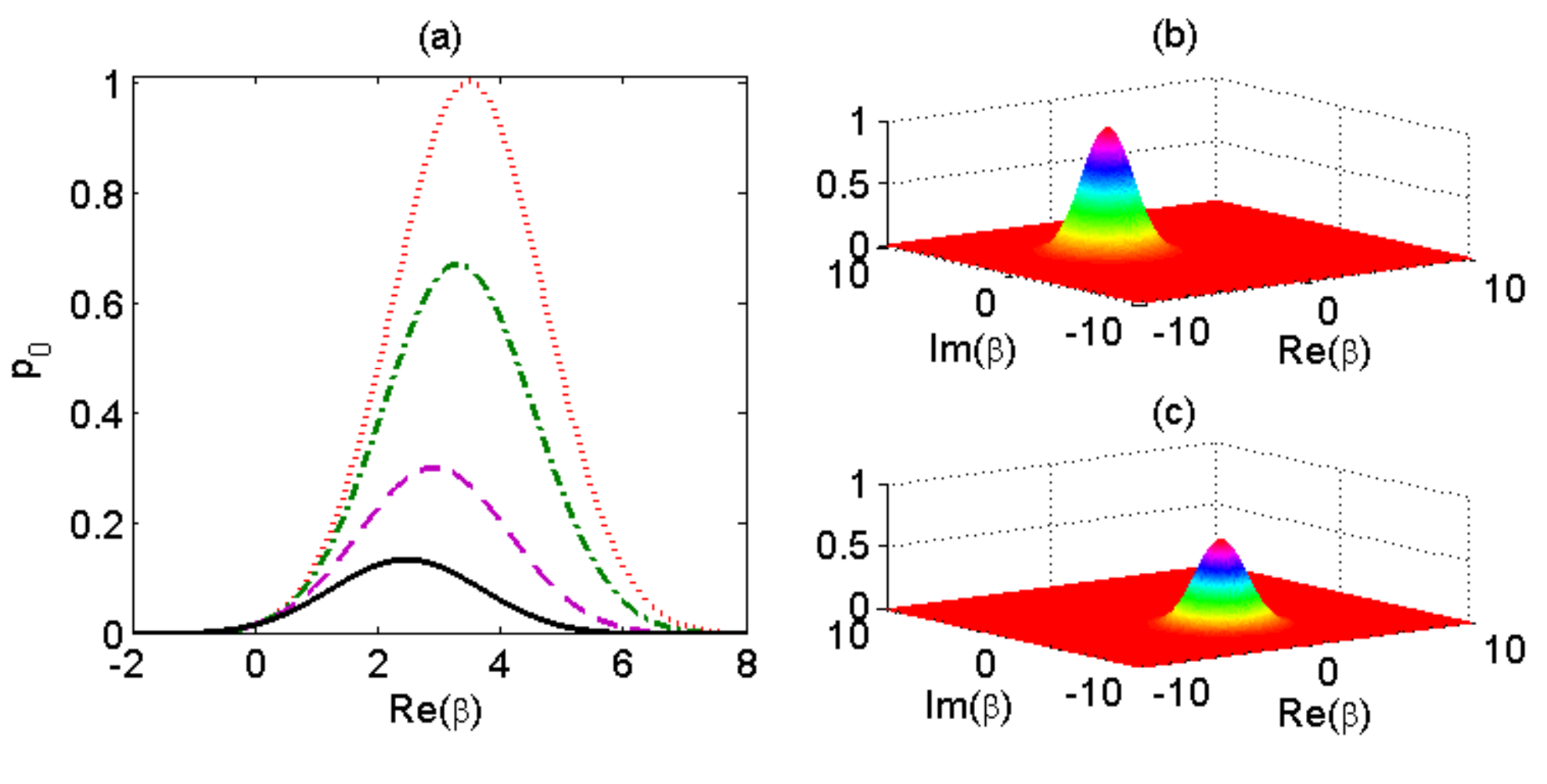}
\caption{{\bf {The zero-click probability distribution as a function of the overlap $\phi$.}}  The panel (a) shows zero-click probability $p_0$ for different overlaps for fixed coherent signal and varied probe. The dotted red, dash-dotted green, dashed purple and solid black lines show $p_0$ for the overlap $|\phi|^2=1,0.9,0.7,0.5$, respectively, and $\arg(\phi)=0$.   The panels (b,c) show zero-click probability for the fixed coherent signal and varied probe, $\beta$, for the two different values of the overlap. The panel (b) corresponds to the intensity overlap equal to 1 and the overlap phase of $3\pi/2$ with respect to the phase between the signal and the probe fields; the amplitude of the signal is 2. The panel (c) is for the overlap equal to 0.95 and the zero overlap phase.  For all the {graphs} the efficiency, $\eta=0.59$; $|U_{1,2}|^2=0.544$; the amplitude of the signal is $2/\sqrt{\eta}|U_{1,1}|$; $U_{1,2}$ and $U_{1,1}$ are taken to be real. }
\label{fig2}
\end{figure}

\section{IV. Inferring the overlap and temporal profile of the signal}
For our set-up, the knowledge trade-off between the
quantum state of the signal and
the spectral shape of the signal TM is realized through the estimation of
the overlap.  In Ref.~\cite{mogilevtsev2009} is was shown
it is possible to infer the modulus of the overlap even for an unknown signal through the implementation
of a  sufficiently large set of coherent probes, so that it is large enough for the
reconstruction of the Wigner function with acceptable accuracy.  However, it is possible to reconstruct
the complete temporal/spectral profile of the signal TM using much smaller set of the coherent states (probes) if a
simple time-delay arrangement is used.

The estimation of the overlap value is based on the analysis of the zero-click probability dependence on the value of the overlap.
Eq.~(\ref{represent}) shows that {altering the} amplitude or phase of the overlap leads to changes {in} the zero-click probability distribution.
For the signal and probe in coherent states this is illustrated in Fig.~\ref{fig2}. Lowering the absolute value of the overlap and changing its phase leads to shifting and damping of the zero-click probability, $p_0(\phi,\beta)$.

We suggest the following general strategy to estimate the overlap:
\begin{enumerate}
\item measure $p_0(\phi,\beta_l)$ for a set of the amplitudes $\beta_l$ and fixed overlap;
\item calculate  $p_0(\phi,\beta_l)$ for different assumed values of the overlap aiming at fitting the measured values.
\end{enumerate}

Notice that for an arbitrary diagonal state (probe or signal), just one value of the probe amplitude is sufficient. Indeed, from Eq.~(\ref{represent}) it follows
that for the diagonal signal $p_0(\phi_1,\beta)\geq p_0(\phi_2,\beta)$ for $|\phi_1|\geq|\phi_2|$. Also, for the non-vacuum signal $p_0(\phi,\beta)$ is strictly increasing with $|\phi|$.

For the set of known overlap values, it is straightforward  to infer the product of spectral envelopes of the signal and probe modes {(for the experimental results see next section)}.  If we vary the overlap using the time-delay for one of the pulses, the product of the frequency profiles of the modes of the signal state, $f_s(\omega)$, and the probe state, $f_p(\omega)${,} is connected to the overlap by the Fourier transform:
\begin{equation}
\phi(\delta t)=\int d\omega f_s(\omega)f_p^*(\omega)\exp{\{-i\omega\delta t\}},
\label{four}
\end{equation}
where $\delta t$ is the value of time-delay.
The problem of the profile inference remains feasible even in the case of when only the modulus of the overlap is estimated. It is a particular case of  much discussed phase-retrieval problem commonly encountered in the image reconstruction (see, for example, the recent brief review \cite{phase}). One of the simplest way{s} to do the profile inference is to represent $f_s(\omega)f_p^*(\omega)$ as a superposition of some localized basis functions, typically the Hermite-Gaussian functions are used. Such a representation is quite useful, for example, for the reconstruction of the states generated in a spontaneous down conversion schemes \cite{ansari}.

\begin{figure}[htb]
\includegraphics[width=0.85\linewidth]{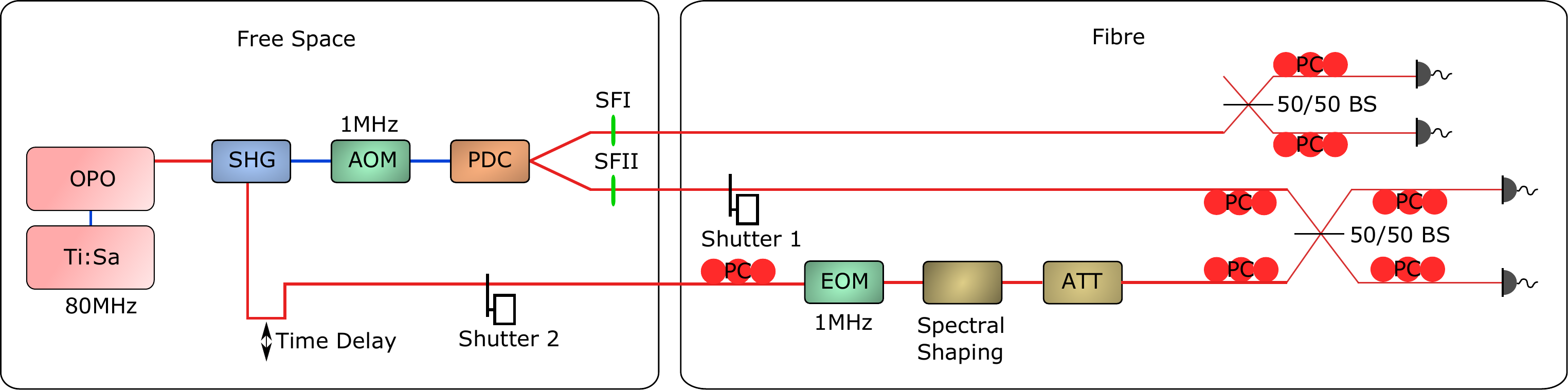}
\caption{{\bf Schematic picture showing the experimental setup.} SF: spectral filter;
EOM: electro-optic modulator; AOM: acousto-optic modulator;
Att: attenuator;
PC: polarisation controller.}
\label{fig3}
\end{figure}

\begin{figure}[htb]
\includegraphics[width=0.75\linewidth]{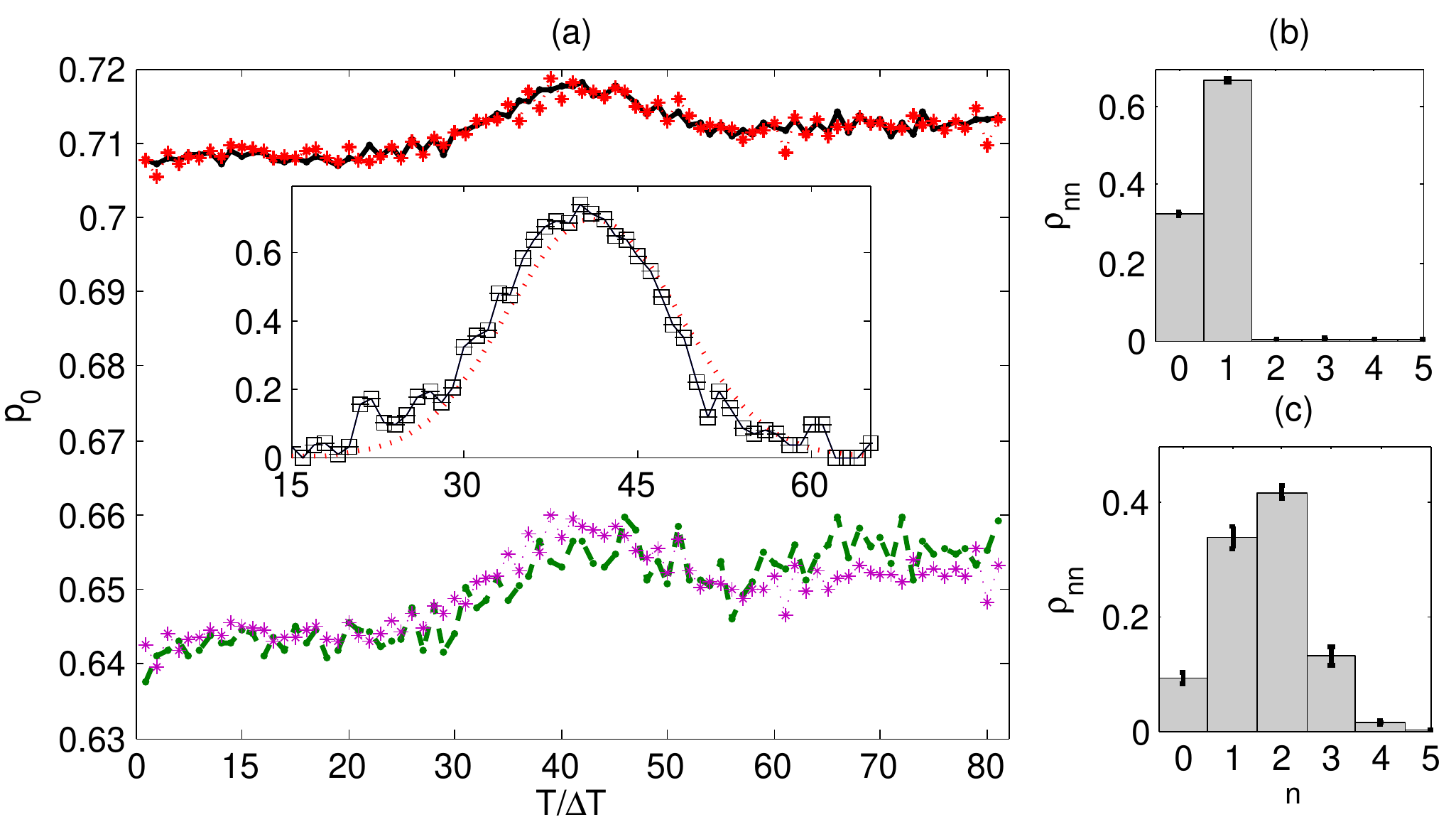}
\caption{{\bf Experimental results for the quantum state reconstruction of the single-mode signal and probe.} (a) Upper curve: Zero-click probability for the single-photon state. Solid black line corresponds to the measured data; dotted red line corresponds to the fit. Lower curve: Zero-click probability for the two-photon state.  Dashed green line corresponds to the measured data; dash-dotted purple line corresponds to the fit. The delay time defining the overlap is $T$, $\Delta T$ is the delay time interval between each $j$-th and $j+1$ measurements; for our the set-up  $\Delta T\approx 33fs$.
The inset shows illustration of the overlap inference for the coherent probe and the single-photon signal.  Dotted red line shows the provided overlap values;   squares shows the inferred overlap values for the single photon state, $0.335|0\rangle\langle0|+0.665|1\rangle\langle1|$.
(b) The inferred photon-number distribution for the single-photon state.  (c) The inferred photon-number distribution for the two-photon state. For all the {graphs} the efficiency, $\eta=0.59$; $|U_{1,2}|^2=0.544$; $U_{1,2}$ and $U_{1,1}$ are taken to be real.}
\label{fig4}
\end{figure}

\section{V. Experimental results and inference}
The setup used for this experiment is shown in Figure~\ref{fig3}. The signal is generated in an engineered periodically poled potassium titanyl phosphate (KTP) waveguide with a type-II parametric down-conversion process \cite{source}. The source can generate a nearly spectrally decorrelated state at 1550 nm. The down conversion is pumped at 775 nm with frequency doubled light from an OPO system. The residual pump light from the OPO at 1550 nm is used as a weak probe field. To adapt the spectrum of the probe field to the signal, spectral shaping involving bandwidth and spectral phase is employed. An automated free space variable delay line is used to change the relative timing between signal and probe. Both fields are then combined on a fibre integrated beam splitter. The idler mode from the parametric down conversion is split on a second 50/50 beam splitter allowing for heralding on one or two clicks.
The light is detected with superconducting nano-wire detectors with a detection efficiency of 90\%.  The repetition rate of the experiment is limited to 1 MHz. An acousto-optic modulator is used to lower the repetition rate for the 775 nm light whereas a fibre integrated electro-optic modulator is used for the probe field.

A total of 81 different delays between signal and probe where scanned with an acquisition time of 150 seconds each. Between these measurements automated shutters where used to either measure the PDC signal or the probe field to account for long term drifts. While only using the PDC signal (shutter 2 is closed) the Klyshko efficiency can be measured giving the product of all efficiencies after the PDC \cite{klyshko1980use}. In addition a calibrated coherent state is used to measure the efficiency from the beam splitter (where signal and probe are combined) to the detectors. Drifts and efficiencies are incorporated in the fits shown in Fig.~\ref{fig4}. The beam splitter has a slight asymmetry with a transmittivity of 43.6.
The overlap is estimated by analysing the Hong-Ou-Mandel interference pattern between probe and signal. The multi-photon components in the signal and probe reduced the visibility of the Hong-Ou-Mandel interference, even in the case of perfect modal overlap.

The results of the quantum state reconstruction for the single-photon and two-photon states are displayed in Fig.~\ref{fig4}. Estimation of the photon-number distributions was performed using Eq.(\ref{represent}) for least squares fit with linear constraints \cite{errorbars}.  The crucial part of the procedure is the experimental inference of the actual value of the overlap as was described in Sec.~IV. The overlap profile inferred for the single-photon states is shown {in the inset of Fig.~\ref{fig4}a.}

Here one should emphasize the different role played by the overlap in different measurement schemes, in particular, in a common homodyning with a strong probe and non-balanced scheme with the weak probe as the one described and implemented in our work. In balanced homodyning with a strong probe the role of the finite overlap is equivalent to additional detection loss. From Eq. (\ref{E91}) it is seen that the role of the overlap for the non-balanced scheme with a weak  probe is quite different.

First of all, the phase of the overlap is rather important (this one can see, for example, in Fig.~\ref{fig2}). As we have demonstrated here, it is possible to build a set of measurements sufficient for the complete reconstruction of the  signal by varying the phase and the modulus of the overlap for the fixed coherent probe. Then, the value of the overlap comes into the equation for the zero-click probability in a different way as the detector efficiency. In fact, the overlap additionally damps and rotates the part of the probe coherent state interfering with the signal, whereas the signal is only damped by the imperfect efficiency (see Eqs.(\ref{E91},~\ref{represent})).  Zero overlap is not the same as zero efficiency for our measurement scheme. Zero efficiency gives unit zero-click probability irrespectively of other parameters. Zero overlap ``factorizes'' a zero-click probability, which for a zero overlap becomes a product of probabilities of probe and signal independently interfering with the vacuum. Notice that for the case of no phase correlation between the signal and the probe, when only the diagonal elements can be inferred, the overlap phase is washed out just like the phase of the probe.

Finally, with the weak probe one can infer the modulus of the overlap even without knowing the signal \cite{mogilevtsev2009}. So, the non-balanced scheme with the weak probe indeed gives the possibility to detect appearance of distinguishability between the signal and probe modes.

\section{Conclusions}

Here we have shown that the state of an arbitrary multi-mode field can be inferred using the imperfect overlap between the signal and the reference fields using just one on/off detector. Interference of the  signal with the controlled probe gives possibility to translate the state of the signal into merely a chain of binary numbers (via the zero-click probability, i.e. probability that the detector does not click). Moreover,
information about the temporal structure of the field can be traded for the information about the
quantum state of this field, and vice versa and reconstruction of both can be realized with the same  measurement set-up. The key for such a measurement scheme is to vary the imperfect mode overlap in a controlled way.

For the  single mode signal and reference fields, we show experimentally that this can be achieved by the simple time delay of the probe.  An important step in the procedure is the experimental estimation of the actual value of the mode overlap $\phi_1$ (\ref{overlap}).
Remarkably, for the single-mode case and for probe or signal in a diagonal state, for the known signal the overlap can be inferred with just one fixed probe state.  The experimental results for the  quantum state inference from the knowledge of the spectral profile by changing the overlap, and the reverse problem of estimating the overlap from the knowledge of the signal state, are depicted in  Fig.~\ref{fig4}.
The suggested scheme can be used for diagnostics of devices (such as quantum pulse gate) which can possibly alter both temporal profile and quantum states of the field.

Thus we have shown that the distinguishability of modes can be a valuable resource. It can be implemented for quantum state diagnostics and tomography. Actually, distinguishability and imperfect overlap of the probe and signal is a bridge which allows to connect spacial, temporal and spectral features of wave-package carrying the quantum state and parameters of this state.

V.S. acknowledge support from the National Council for Scientific and Technological Development (CNPq) of Brazil,  grant  304129/2015-1, and by  the S{\~a}o Paulo Research Foundation   (FAPESP), grant 2015/23296-8. D.M.  acknowledge support from the EU
project Horizon-2020 SUPERTWIN id.686731,
the National Academy of Sciences of Belarus program
"Convergence" and FAPESP grant 2014/21188-0. N. K. acknowledges
the support from the Scottish Universities Physics Alliance (SUPA) and  from the International Max
Planck Partnership (IMPP) with Scottish Universities.   J.T. and C.S. acknowledge support from European Union Grant No.665148 
(QCUMbER). T.B. acknowledges support from the DFG under TRR 142. 

\section{Appendix}

The multidimensional  Weierstrass transform of a function $Q(\alphab)$ of a complex variable $\alphab = (\alpha_1,\ldots,\alpha_S) \in C^{S}$ is defined as a convolution with the   following multidimensional Gaussian  (for $\sigma>0$)
\be
\mathcal{W}(\alphab,\alphab^\dag) = \int d\mu(\betab) e^{-\sigma|\alphab-\betab|^2}Q(\betab,\betab^\dag),
\en{E16}
where  $\betab = (\beta_1,\ldots,\beta_S)\in C^S$.  Assuming that there is the  (multidimensional) Fourier transform of $Q(\alphab)$ (as in the case of the  Husimi function Eq. (\ref{E7})), one can show that  there is the inverse transform to Eq. (\ref{E16}). Let us write the Fourier transform using the complex variables:
\be
Q(\alphab,\alphab^\dag)= \int \frac{d\mu(\zb)}{\pi^S} \exp\left\{\zb\alphab^\dag - \alphab\zb^\dag \right\}F(\zb,\zb^\dag),\quad
F(\zb,\zb^\dag) = \int \frac{d\mu(\alphab)}{\pi^S} \exp\left\{ \alphab\zb^\dag - \zb\alphab^\dag\right\}Q(\alphab,\alphab^\dag).
\en{E17}
Substituting the Fourier transform of Eq. (\ref{E17}) into Eq. (\ref{E16}) and evaluating the Gaussian integral with the help of  Eq. (\ref{E10}), we can put the Weierstrass transform in the differential operator form (i.e.,  an infinite series expansion):
\be
\mathcal{W}(\alphab,\alphab^\dag) = \left(\frac{1}{\sigma}\right)^S\int d\mu(\zb)\exp\left\{\zb\alphab^\dag - \alphab\zb^\dag - \frac{|\zb|^2 }{\sigma}\right\}F(\zb,\zb^\dag)=\left(\frac{\pi}{\sigma}\right)^S\exp\left\{\frac{1}{\sigma}\sum_{s=1}^S\frac{\partial^2}{\partial \alpha_s^*\partial \alpha_s}\right\}Q(\alphab,\alphab^\dag).
\en{E18}
Eq. (\ref{E18})  has the inverse transformation (in the form of an infinite series)
\be
Q(\alphab,\alphab^\dag) = \left(\frac{\sigma}{\pi}\right)^S\exp\left\{-\frac{1}{\sigma}\sum_{s=1}^S\frac{\partial^2}{\partial \alpha_s^*\partial \alpha_s}\right\}\mathcal{W}(\alphab,\alphab^\dag).
\en{E19}
The infinite series representation of Eq. (\ref{E19}) for the Husimi function from its Weierstrass transform can be recast in the explicit form as  an  integral in $R^{2S}$. Substituting the operator identity
\be
 \exp\left\{-\frac{1}{\sigma}\sum_{s=1}^S\frac{\partial^2}{\partial \alpha_s^*\partial \alpha_s}\right\} = \left(\frac{\sigma}{\pi}\right)^{S}\int d\mu(\zb) \exp\left\{ -\sigma|\zb|^2 +i\left(\sum_{s=1}^S z^*_s\frac{\partial}{\partial \alpha_s} +z_s\frac{\partial}{\partial \alpha^*_s}\right)  \right\}
\en{E20}
(derived by using Eq. (\ref{E10}))  into Eq. (\ref{E19}) and  setting $z_s = x_s +iy_s$, we obtain:
\begin{eqnarray}
Q(\alphab,\alphab^\dag) &= & \left(\frac{\sigma}{\pi}\right)^{2S}\int d\mu(\zb) \exp\left\{ -\sigma|\zb|^2 +i\left(\sum_{s=1}^S x_s\frac{\partial}{\partial \mathrm{Re}\alpha_s} +y_s\frac{\partial}{\partial \mathrm{Im}\alpha_s}\right)  \right\}\mathcal{W}(\alphab,\alphab^\dag)\nonumber\\
&=& \left(\frac{\sigma}{\pi}\right)^{2S}\int \prod_{s=1}^Sdx_sdy_s  \exp\left\{ -\sigma\sum_{s=1}^S(x_s^2+y_s^2)  \right\}\mathcal{W}_R(\mathrm{Re}\alphab+i\mathbf{x},\mathrm{Im}\alphab + i\mathbf{y}),
\label{E201}\end{eqnarray}
where $\mathcal{W}_R(\mathrm{Re}\alphab,\mathrm{Im}\alphab) $ is the Weierstrass transform $\mathcal{W}(\alphab,\alphab^\dag)$ expressed as a function of the real and imaginary parts of the complex vector $\alphab$.


\begin{thebibliography}{99}
\bibitem{errorbars} Error estimation was performed for 100 sets of the input data generated by randomly taking out one value from 15 of those corresponding to each overlap value and summing the rest; for each set the second bootstrap was performed taking 1000 random sets of 81 overlap points.
\end{thebibliography}
\end{document}